\documentclass[sigconf, authorversion, nonacm]{acmart}

\AtBeginDocument{%
  \providecommand\BibTeX{{%
    \normalfont B\kern-0.5em{\scshape i\kern-0.25em b}\kern-0.8em\TeX}}}

\usepackage{caption}
\usepackage{subcaption}

\setcopyright{acmlicensed}
\copyrightyear{2018}
\acmYear{2018}
\acmDOI{XXXXXXX.XXXXXXX}

\acmConference[Conference acronym 'XX]{Make sure to enter the correct
  conference title from your rights confirmation emai}{June 03--05,
  2018}{Woodstock, NY}
\acmBooktitle{Woodstock '18: ACM Symposium on Neural Gaze Detection,
 June 03--05, 2018, Woodstock, NY} 
\acmISBN{978-1-4503-XXXX-X/18/06}

\begin{document}

\title{Acceptance of an Augmented Society: Initial Explorations into the Acceptability of Augmenting Real World Locations}

% Acceptance of the Augmented Society: Initial Explorations into the Acceptability of Augmenting Public, Private, Religious and Cultural Real World Locations


\author{Alan Joy}
\affiliation{%
  \institution{University of Glasgow}
  \city{Glasgow}
  \country{Scotland}}
\email{2460256j@student.gla.ac.uk}

\author{Joseph O'Hagan}
\authornote{Corresponding author}
\affiliation{
  \institution{University of Glasgow}
  \country{Glasgow, Scotland}
  }
\email{joseph.ohagan@glasgow.ac.uk}

\renewcommand{\shortauthors}{Joy and O'Hagan}

\begin{abstract}
Augmented reality (AR) will enable individuals to share and experience content augmented at real world locations with ease. 
But what protections and restrictions should be in place? 
Should, for example, anyone be able to post any content they wish at a place of religious or cultural significance? 
% We investigated attitudes towards the posting a variety of AR content types at different real world locations. 
We developed a smartphone app to give individuals hands-on experience posting and sharing AR content. 
After using our app, we investigated their attitudes towards posting different types of AR content (of an artistic, protest, social, informative, and commercial nature) in a variety of locations (cultural sites, religious sites, residential areas, public spaces, government buildings, and tourist points of interests). 
Our results show individuals expect restrictions to be in place to control who can post AR content at some locations, in particular those of religious and cultural significance. 
We also report individuals prefer augmentations to fit contextually within the environment they are posted, and expect the posting and sharing of AR content to adhere to the same regulations/legislation as social media platforms. 
\end{abstract}


% \ccsdesc[500]{Information systems~Sentiment analysis}


\begin{teaserfigure}
  \centering
  \includegraphics[width=0.55\textwidth]{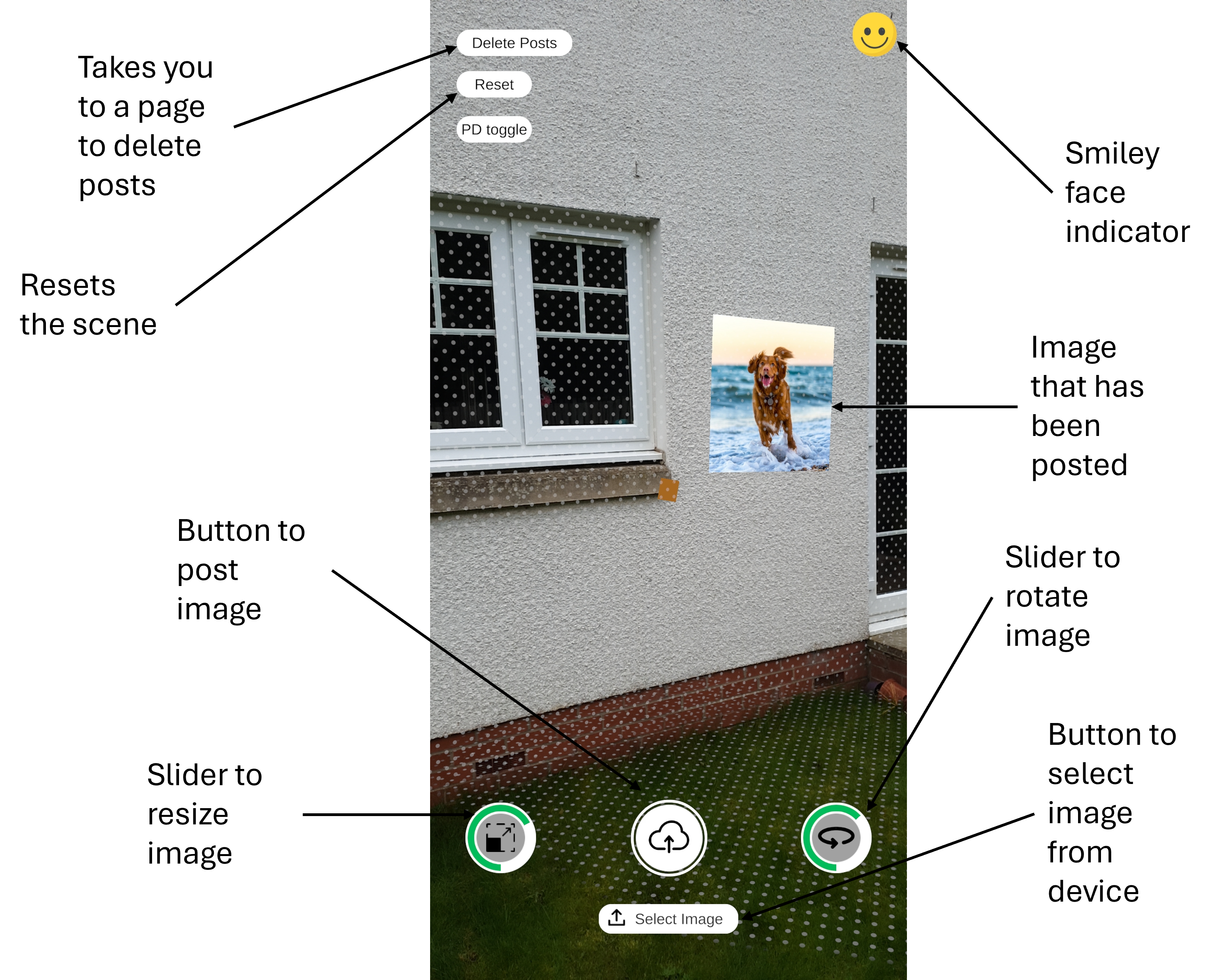}
  \caption{A demonstration of our smartphone app. All of the functionality of the app is labeled. A photo of a dog as AR content is attached to the side of a house. The smiley face indicator informs the user of the strength of the feature map where a smiley face means it is of sufficient quality to post content.} 
  \Description{A demonstration of our smartphone app. All of the functionality of the app is labeled. A photo of a dog as AR content is attached to the side of one of the researchers' houses. The smiley face indicator informs the user of the strength of the feature map where a smiley face means it is of sufficient quality to post content.}
  \label{fig:teaser}
\end{teaserfigure}


%   \includegraphics[width=\textwidth]{misc_files/sampleteaser.pdf}


\maketitle

%%
%% I HAVE INCLUDED SECTION TEX IN tex_files/file.tex FOR EACH SECTION TO TRY AND MAKE THIS A BIT MORE MANAGEABLE - IT PROBABLY WON'T BE
%%

\section{Introduction}
As augmented reality (AR) devices tend towards everyday/all-day, ubiquitous and fashionable form factors \cite{imwut-joseph, mmve-joseph, mathis2023breaking} (e.g. smart glasses \cite{imwut-joseph}), users will be able to share and experience AR content interwoven with people \cite{jolie-vrst} and the physical world that surrounds them \cite{intro_3}.
For example, a city street could be filled with AR artwork \cite{street-art}, historical information could be placed next to points of historical significance \cite{intro_6}, and more \cite{mum-essay}. 
However, this technology may also be used more controversially. 
Individuals might be harassed by others who attach derogatory AR messages to their home \cite{intro_7}, or businesses could be defaced with AR graphics posing a risk to their professional image and reputation \cite{intro_8}. 
And individuals will be able to do this with ease, creating and disseminating harmful AR content by simply directing a camera/sensor and pressing a few buttons. 
Users can already do this using existing social platforms like Snapchat or TikTok, but in the near future might contribute their posts to shared AR world views \cite{dig-graf-1, dig-graf-2}, increasing the content's reach and potentially even its impact. 

Such capabilities evoke clear concerns surrounding the use of such potential technologies. 
Is it, for example, acceptable for a religious site (e.g. a church) to be augmented with commercial content, or even content posted by the general public? 
Or should some locations be protected so only approved parties may post there?
While recently discussion into such issues has increased (e.g. \cite{lit_1, lit_2, lit_3}) further research is needed. 

To address this, we contribute a user study (N=12) which investigated attitudes towards the augmentation of common, everyday, real world locations (cultural, religious, public, residential, government, and tourist locations) with a variety types of AR content (artistic, protest, social, informative, and commercial content). 
In our study, participants gained hands-on experience using a smartphone app we developed to post/view persistent AR content (2D images) then completed a questionnaire and interview to reflect on their attitudes towards the acceptability of augmenting real world locations. 

We show participants believed the creation, sharing, and viewing of AR content in real world locations should be regulated/legislated in the same way existing social media platforms regulate/legislate user content. 
Additionally, participants expect restrictions to be in place to control who can post AR content at some locations, in particular those of religious and cultural significance. 
When augmenting locations you own, participants expect to be free to do as they wish provided the augmentations do not break any existing laws around free speech. 
Participants also expect AR advertisements to adhere to existing standards, i.e. being located where existing physical advertisements already are and being age and content appropriate for their setting. 

\section{Implementation}
\label{implementation}
We developed a smartphone app to allow users to post and view persistent AR content in real world locations. 
We used Unity and ARCore for development. 
A Pixel 8 was used to run our user study.

\subsection{Application Functionality}
Our app was designed to enable users to post a local image from a smartphone as AR content in a user's surrounding environment, attaching the image to either a vertical or horizontal surface (e.g. a floor, wall, table, etc).
For the study, the smartphone was setup with a selection of creative commons images including graffiti styled pop art and photos of dogs and cats.

After selecting an image, users could attach the image to a targeted surface and adjust its position and placement using sliders. 
When satisfied, they could then upload the image as persistent content in an AR view of the environment. 
To upload the image, users were required to walk around the placed image in a small semi-circle (keeping the camera fixed on it) until a notification indicated the post was fully captured and hosted.
This process was used to increase the accuracy of the underlying feature map used to attach the image as AR content. 
Any posted image would persist at the attached location in the AR view of the environment across use sessions, and would remain until deleted by a user using a menu in the app. 
Figure \ref{fig:teaser} shows a screenshot of the app in use.

\subsection{Implementation Details}
When the camera identified a plane to attach an image onto, an orange placement indicator appeared to inform the user they could place an image at this location. 
The user could then place an image by selecting it from their device using the \textit{``Select Image''} button and tapping anywhere on the screen. 
A sheet of faint dots appeared across detected planes to show detected surfaces where content could be placed.
To achieve image persistence we used ARCore’s Cloud Anchors \cite{design_3} to enable the persistent sharing of AR content across sessions and devices. 
When the user places an image, ARCore constructs a 3D feature map of the environment and assigns a unique Cloud Anchor ID to the posted image that was then stored on Google Cloud (Firebase Realtime Database \cite{design_7}). 
This ID and 3D feature map can then be utilised later to resolve the anchor to re-instantiate the image at the correct position in the environment and create content persistence across devices and sessions \cite{design_6}. 

\section{Study Design}
We conducted a user study where participants gained hands-on experience using our app and then completed a questionnaire and interview to explore their attitudes towards augmenting real-world locations with AR content. 
We opted for this approach to ensure participants had hands-on experience with how such a future system might work to better inform their answers.

\subsection{Procedure}
Upon arrival the experiment’s purpose was explained and a consent form and demographic questionnaire given to the participant.
Participants were told they would first gain hands-on experience with an app designed to demonstrate how future AR content could be posted and shared.
After this, participants were told we would investigate their attitudes towards the acceptability of augmenting different real-world locations with AR content using a questionnaire and interview approach. 

The experiment took place in a student computer lab for undergraduate and masters students on a university campus. 
This environment was chosen as a semi-public space, and an environment in which the app was thoroughly tested to work (enabling placement of AR content on the many different types of horizontal and vertical surface within the room). 
Conducting the experiment here also avoided differences in weather and lighting conditions across participants, ensuring a more consistent experience using our app across participants. 
Although, we acknowledge this as a limitation of our work. 

The experiment took approximately 45 minutes to complete, consisting of a: 5-minute introduction (explaining augmented reality generally, the experiment's purpose, and how the application worked), 15-minute session to use the app to complete a set of tasks, and 25-minute session to completed a questionnaire and semi-structured interview. 
During the 15-minute hands-on session, participants were encouraged to think aloud and explain actions they were taking or any thoughts they had about the app generally. 

Although AR content placed during the experiment by a participant could persist across sessions (e.g. across different participants) we reset the application to a baseline augmentation state after each participant completed the experiment. 
This removed the prior participant's AR content but left some placed by the researcher in the environment. 
This was done to ensure a comparable starting experience for all participants in our study. 
This baseline state included three examples of posted AR content placed next to where participants were instructed to stand when starting the study.

\subsection{Experimental Task}
\label{tasks_outline}
During the hands-on session participants were tasked with completing the following: 

\begin{enumerate}
    \item \textit{Open the app, and locate the three example images hosted in the surrounding environment.} 

    \item \textit{Post an image within the environment and make it persistent: Select an image from device and scan the environment to identify a surface to place the AR content on. Use the placement indicator to place image at desired location. Use the scale and rotation sliders to adjust size and orientation of image as desired. Scan the environment and click the host button.}
    
    \item \textit{Resolve the hosted image and verify its persistence: Close and reopen the app. Locate the image you just hosted.}
    
    \item \textit{Spend the remaining time as you like, posting images wherever you want in the environment, exploring the features and capabilities of the application.}
\end{enumerate}

\subsection{Questionnaire Design}
Our questionnaire asked participants to rate the acceptability, on a 5-point Likert scale (1 = Extremely Unacceptable, 5 = Extremely Acceptable), of five types of AR content (\textit{Artistic, Protest, Social, Informative, Commercial}) being augmented in six locations (\textit{Cultural Site, Religious Site, Residential Area, Public Space, Government Building, Tourist Point of Interest}). 
The questionnaire was structured so that participants were presented with a location and then asked to rate the acceptability of posting each of the five types of AR content at this location. 
The location order was counterbalanced using a fully balanced Latin square. 
The order of the AR content types within each location was randomised. 
Participants were instructed to assume the individual posting the AR content did not have ownership over the space being augmented and that the content would be posted on a public, shared, AR view of the location.  

Each location was presented with a representative graphic and short text description to aid participants' understanding of the location (Appendix \ref{app:appendix_A}). 
Each AR content type was presented with a text description. 
This was done to mitigate the potential influence of an example augmentation being the focus of participants' reflection, rather than content type more generally. 
Due to the visual only focus of our app, our text descriptions focused on visual only augmentations as well. 
The text descriptions used to describe each content type follow: 

\begin{itemize}
    \item \textit{\underline{Artistic}}: refers to creative expression, e.g. visuals and artworks aimed at conveying ideas, emotions, or aesthetics.
    
    \item \textit{\underline{Protest}}: refers to any imagery or text created that conveys dissent, criticism, or advocacy for social or political causes.
    
    \item \textit{\underline{Social}}: refers to messages exchanged between friends or images intended for sharing with others, reflecting personal experiences, opinions, etc.
    
    \item \textit{\underline{Informative}}: refers to data or visuals that provide relevant and useful information about the location or surrounding environment.
    
    \item \textit{\underline{Commercial}}: refers to advertisements, promotions, or marketing materials aimed at promoting products, services, or brands for financial gain.
\end{itemize} 

We selected a list of locations that are common in everyday life and have been shown in prior works to be locations individuals are willing to augment with AR content in some capacity \cite{dig-graf-1, intro_6, imx-church-paper}. 
We selected our AR content types based on common uses cases of AR (e.g. social messages to friends \cite{dig-graf-1}, street art \cite{street-art}, protest art \cite{protestAR}, etc) and on common physical alterations made by individuals.

\subsection{Interview Questions}
After completing the questionnaire, a semi-structured interview was conducted to allow participants to elaborate on the reasoning behind their answers, and to explore their attitudes towards protecting locations from AR content. 
The questions asked to all participants were: 
\begin{itemize}
    \item \textbf{IQ1:} \textit{``If an individual owns a location, do you believe they should have the unrestricted right to augment it as they please?'' }

    \item \textbf{IQ2:} \textit{``Are there any specific limits or guidelines you would propose to regulate augmentations on private locations?''}
 
    \item \textbf{IQ3:} \textit{``For any of the locations you saw in the questionnaire, do you believe you should be required to have explicit permission from the owner in order to augment it with AR content? Explain your reasoning.''}

    \item \textbf{IQ4:} \textit{``Do you think there should be particular regulations concerning any of the AR content categories you saw in the questionnaire, and if yes, what should they entail?''}
\end{itemize}

\begin{figure*} [h]
\begin{center}
\includegraphics[scale=0.4]{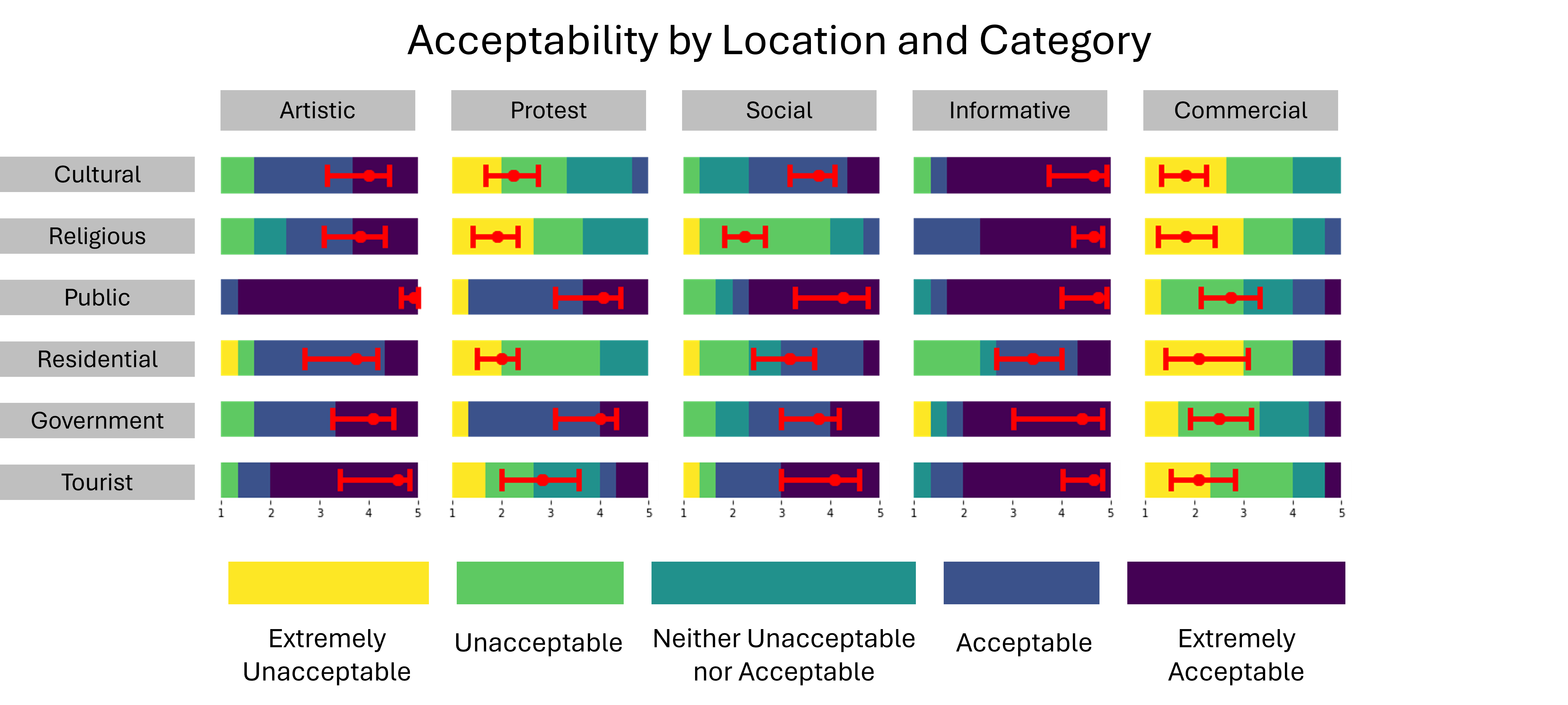}
% \vspace{-15pt}
\end{center}
\caption{Responses to Likert scale questions surveying the acceptability of augmenting different locations with different content types using an AR graffiti app. 95\% confidence intervals are visualized with red bars, based on the conversion of ordinal variables to numeric ranks (1 = Extremely Unacceptable and 5 = Extremely Acceptable)}
\label{fig:ethics-results}
\end{figure*}

\section{Results}

\subsection{Participant Demographic Data}
Participants were recruited using social media and mailing lists. 
Participation in the study was voluntary. 
12 participants completed the study (2 female, 10 male) aged between between 21 and 23 years of age (M=22.1, SD=0.49). 
Participants indicated prior familiarity with smartphone AR (M=3.25, SD=1.01; 5-point Likert scale; 1 = Not At All Familiar; 5 = Very Familiar), and with AR headsets (M=2.42, SD=1.19; 5-point Likert scale; 1 = Not At All Familiar; 5 = Very Familiar). 

Our participants were exclusively based in The University of Glasgow, Scotland.
Consequentially, discussions may be influenced by predominantly Western cultural and societal norms. 
E.g. this could influence participants' opinion on the acceptability of augmenting cultural or religious sites. 
Furthermore, participants were around a similar age range and bias towards being male. 
It is important then to approach the generalisability of our results with this in mind, and future work should look to explore multi-cultural attitudes and a wider demographic range.

\subsection{Analysis}
An Aligned-Rank Transform (ART) \cite{results_1} was used for significance testing to transform non-parametric data prior to conducting a repeated measures ANOVA, using the ARTool R package \cite{results_2}. 
ART enables the use of parametric tests on non-parametric data (e.g. Likert-type responses \cite{likert}). 
When using ART, as noted by Wobbrock et al. \cite{wobbrockART} \textit{``the response variable may be continuous or ordinal, and is not required to be normally distributed''}, making it well suited to our dataset. 
Where two factors of concern existed, a two-way ANOVA was conducted (comparing location with AR content type). 

Figure \ref{fig:ethics-results} shows the distribution of acceptability for each AR content type at each location. 
This plot includes 95\% confidence intervals (visualised with red bars) based on conversion of dependent ordinal variables to numeric ranks, allowing a by-eye estimation of significant pairwise differences (where the confidence intervals do not overlap) - an approach favoured by those who believe reporting should be done with confidence intervals and visualisations \cite{errorbars}.

Participants' qualitative data collected in the interviews was coded using initial coding \cite{openaxebook} where participants’ statements were assigned emergent codes over repeated cycles with the codes grouped using a thematic approach. 
A single coder performed the coding (2 cycles) and reviewed the coding with one other researcher.
A Google Pixel 4a was used to record participants' qualitative data and to transcribe them.

\subsection{Questionnaire Results}
A two-way repeated-measures ANOVA was performed using an Aligned-Rank Transform (ART) with the ARTool R package. 
This revealed a significant effect on location (F=24.83, p < 0.001) and AR content type (F=84.18, p < 0.001). 
It also revealed significant interactions between location and AR content type (F=4.71, p<0.001). 
This suggests there are significant differences in acceptability ratings across different locations and there are significant differences in acceptability ratings across different content types.
It also suggests the effect of location on acceptability may vary depending on the AR content type, and vice versa.
These insights can also be seen by visually inspecting the differences in error bars in Figure \ref{fig:ethics-results}. 

Of note from Figure \ref{fig:ethics-results} is that, generally, \textit{Artistic} content was considered acceptable to be posted in all of our proposed locations. 
In contrast, \textit{Commercial} content was consider mostly unacceptable across all the locations. 
Attitudes towards \textit{Informative}, \textit{Protest} and \textit{Social} content were more location specific. 
\textit{Informative} content was considered by most to be acceptable in all of our locations, although a subset of participants did consider it unacceptable in \textit{Residential} locations. 
\textit{Protest} content was considered acceptable in \textit{Public} and \textit{Government} locations but not in \textit{Residential} or \textit{Religious} locations. 
While \textit{Social} content was considered unacceptable in \textit{Religious} locations but acceptable in all of the others.

\subsection{Insights From The Interviews}
\label{results:interviews}

\subsubsection{\textbf{Augmenting Locations You Own:}} all participants (n=12) agreed individuals should have the right to augment a location they owned provided \textit{``it does not break any existing laws''} and \textit{``the content is not deeply offensive or inappropriate''}.

\subsubsection{\textbf{Augmenting Locations You Do Not Own:}} 
All participants said they would restrict access to posting AR content in religious and culturally significant locations.
Instead, they felt individuals who post content in these locations should have approved permission before doing so. 
Participants were concerned without control over who can post in these locations it would lead to misuse, disrespect, harassment, and harm, e.g. overlaying religious imagery with modern non-religious artwork, spamming inappropriate content contrary to the values/beliefs associated with a location, etc. 
% Participants felt by controlling who can post in these locations (and what) that the potential for harm and misuse mitigated against.

8 participants said they would restrict AR content from being posted in residential areas due to concerns surrounding the cyberbullying or harassment. 
In particular, the participants were concerned for an individual's residence to be targeted because of their own personal beliefs with hateful AR content, e.g. digital equivalents of a cross being burned into an individual's lawn \cite{mum-essay}. 

% 3 participants also stated they would restrict AR content from being posted in government locations - highlighting the often historical and cultural significance of these locations, and value as a whole to their surrounding community. 
Alongside content control, 1 participant discussed the importance of context when considering significant locations. 
This participant discussed changing attitudes towards potentially sensitive historical locations, \textit{“It feels more acceptable to augment the Great Wall of China because its historical significance may not be as immediately relevant today, but it wouldn't be appropriate to augment a Holocaust memorial.”}.
Although this too requires cultural consideration, as Western and Eastern individuals may have very different attitudes to the point raised by this participant.

\subsubsection{\textbf{Regulating and Restricting AR Content Types:}} 
11 participants said there should be significant regulation on commercial AR content. 
9 of these participants said locations where commercial AR content could be placed should be controlled to avoid disrupting individuals (e.g. posting advertisements at historical locations or areas of natural beauty) or be disrespectful (e.g. posting advertisements in religious or memorial locations). 
These participants felt commercial AR content should be: posted only where existing advertisements are, ideally relevant to the location in which it is placed (to increase the value to anyone who sees it), and be age appropriate, e.g. \textit{``people or companies shouldn't be allowed to place advertisements aimed at adults at a children's park''}.

5 participants said there should be significant regulation and control over AR content about protesting.
All indicated this was due to the risk of protest content evoking a harmful, dangerous, or violent reaction. 
As such, these participants considered it important protest content adhere to regulations to mitigate against the risks the content could provoke.

\subsubsection{\textbf{Comparisons to Social Media Platforms:}} all participants stated any app designed to host and/or share AR content should be restricted by the same guidelines as social media platforms and their content standards. 
9 participants directly compared our app to existing social media because of its emphasis on the creation and sharing of content. 
As such, they believed apps like ours should be held to the same restrictions and legislation that existing mainstream social media platforms (e.g. Instagram, Snapchat, X, etc) are, focused on ensuring content is age appropriate, legal, not offensive or hate speech, adhere to expected social norms, etc \cite{results_4}. 

\section{Discussion}
Our results show attitudes towards the acceptability of AR content depends on both the location of augmentation and the type of augmentation being made. 
Public spaces, for example, generally considered shared spaces belonging to the community rather than any individual or organization \cite{disc_1, disc_2} were considered acceptable for posting AR content. 
In contrast, other locations, e.g. those of religious or cultural significance, were considered to be in need of protection, e.g. to only be augmented by those with permission to do so to avoid being augmented with content in a disrespectful or hateful manner.

\subsection{Legislation and Regulation}
In addition to restricting who could post at certain locations, participants often drew similarities between how existing social media platforms (e.g. Snapchat, TikTok, etc) manage content and how shared AR content might be managed in the future. 
That is, participants expected the posting, sharing, and viewing of AR content to follow the same regulations/legislation that posting, sharing, and viewing content on social media platforms currently follow. 
However, future work is necessary to determine if the existing scope of guidelines, restrictions, regulations, and legislation are comprehensive enough to address the platform specific harms which AR content might enable. 

One distinguishing aspect of AR content, compared to social media posts, is that even if the content itself does not breach existing laws, its placement in certain contexts may still result in significant harm. 
Imagine augmenting a synagogue with images depicting wealth or money. 
Even though the content itself, in isolation, might not violate specific hate crime laws, its underlying sentiment and placement in this location, i.e. its context, makes it highly inappropriate and harmful. 
This distinction underscores the need to consider the differences between AR content and existing social media platforms, and what additional measures will be necessary due to the unique harms AR can enable.

\subsection{Suggestions for Future Work}
We close by highlighting directions future research can take to further investigate the acceptability of augmenting real world locations, but also to develop effective protections for locations to mitigate and/or restrict augmentations when necessary.  

Future work is needed to determine what locations should be restricted partially or fully from being augmented by the general public. 
Our work indicates locations of religious or cultural significance, and possibly even residential areas, are potentially sensitive locations where such measures should be taken. 
The need for such restrictions has already been evidenced.
E.g., shortly after the release of Pokemon Go the Auschwitz-Birkenau State Museum was forced to ban individuals from playing the game whilst at the Auschwitz concentration camp \cite{pokemon-go} and asked the game's developer to issue an update making the game unplayable at this location. 

Future work should also aim to be cross-cultural and capture a global perspective to obtain a more comprehensive view of public augmentation use and an understanding of what is considered an unacceptable augmentation. 
Such work could utilise surveys, but could also look for opportunities to conduct remote user studies \cite{flo-alt-methods, flo-alt-methods-2, flo-alt-methods-3}, or even run the same study in multiple countries, e.g. \cite{ammar}.

In addition to considering what legalisation might be needed \cite{mum-essay, essay-short}, future work should also investigate what other protections can be developed.
Privacy enhancing technologies, for example, are designed to protect the rights of people around AR users from potential harms the AR device could cause to them \cite{imwut-joseph}, and equivalent protection systems could be developed to protect locations, places, and spaces from augmentation. 
One solution might be to re-introduce physical restrictions, e.g. requiring a user to physically touch a surface before allowing them to augment it with content \cite{sphere, van2004tangible, billinghurst2008tangible}. 
Such an approach would introduce a physical deterrent to posting AR content but also severely restrict its potential use. 

From a use standpoint, future work is also needed to understand how users will engage with the augmented content we have discussed throughout this paper. 
One can easily envision on a public, shared AR world view that some locations will contain large amounts of content. 
The Sekia Camera app \cite{lit_9}, which provided users with a \textit{``real-time, location-based, augmented reality social networking system''}, encountered this issue where a large numbers of posts were made in popular locations, overwhelming users screens with overlapping content, breaking the app's functionality. 
To address this, the developers allowed users to manually search for or filter posts at a given location. 
But how could such a system automatically resolve this problem for users? 
Suppose, for example, five overlapping pieces of AR content are applied to a building a user is looking at. 
The user's device could utilise the spatial flexibility \cite{hyunsung-minexr, o2023dynamic, chi-joseph} and inherent dynamic presentation \cite{imx-joseph, avi-joseph,  bystander-perdis-joseph, shady-bans, julie, ismar-stories} of XR content to contextually alter the position of overlapping content to optimise clarity and visibility for the user \cite{reality-aware, ieee-hyunsung}. 
Or a recommendation system could algorithmically decide which content to prioritise and show a user, based on the content they are mostly likely to engage with.

Finally, as individuals begin to post, share, and view AR content in real world locations more widely, it is worth monitoring discussion of this on social media platforms. 
User posts and discourse on social platforms are increasingly becoming a valuable data source for HCI researchers \cite{singh2024exploring, user-reviews, subreddit-addition-recovery-gauthier, ux-subreddit-shukla} as a way to understand sentiment towards new technology \cite{li2023sentiment} or its use \cite{ux-subreddit-shukla}. 
Monitoring emergent trends and discourse on such platforms will, eventually, provide insight into public attitudes on what is considered an acceptable augmentation or not, reported instances where protective measures have failed, and more.

\section{Conclusion}
We present initial insights into attitudes towards the acceptability of augmenting real world locations with AR content. 
We conducted a user study (N=12) where participants gained hands-on experience posting/viewing AR content using a smartphone app we developed. 
After this, participants then completed a questionnaire and interview to reflect on their attitudes towards the acceptability of augmenting cultural sites, religious sites, residential areas, public spaces, government buildings, and tourist points of interests, with AR content of an artistic, protest, social, informative, and commercial nature. 
While more work is needed, we provide early insights into the expectations of individuals. 
We show participants all expected the creation, sharing, and viewing of AR content to be regulated/legislated in the same manner as content posted on existing social media platforms. 
Additionally, they expected some locations, e.g. those of religious and cultural significance, to be restricted so that only approved individuals were permitted to augment. 
When augmenting locations you own, participants expected to be free to augment as they wish provided no existing laws are broken. 
And concerning AR advertisements, participants said believed these should mimic how existing real world, physical advertisements are embedded into locations.


% \begin{acks}
% Nothing
% \end{acks}

%% The next two lines define the bibliography style to be used, and the bibliography file.
\bibliographystyle{ACM-Reference-Format}
\bibliography{references}

\newpage
\onecolumn % Switch to one-column layout
\appendix

\section{Location Images used in our Questionnaire}
\label{app:appendix_A}
\begin{figure}[h]
  \centering
  \begin{minipage}{0.3\textwidth}
    \centering
    \includegraphics[width=\linewidth]{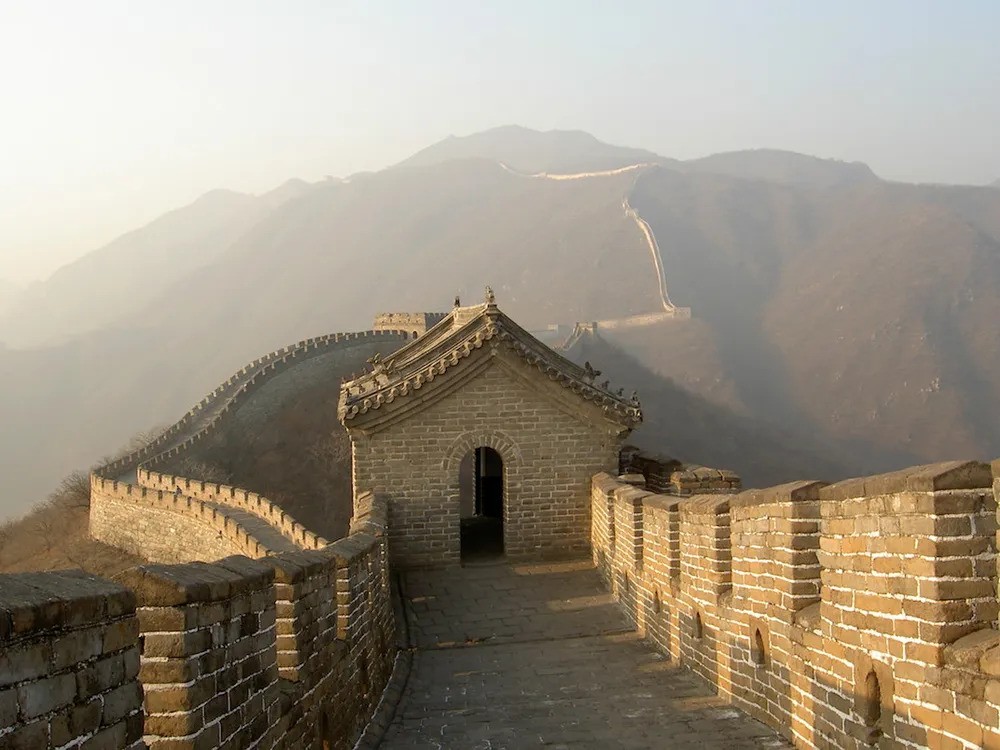} % Replace 'example-image-a' with the filename of your first image
    \caption{Cultural Site}
    \label{fig:first}
  \end{minipage}\hfill
  \begin{minipage}{0.3\textwidth}
    \centering
    \includegraphics[width=\linewidth]{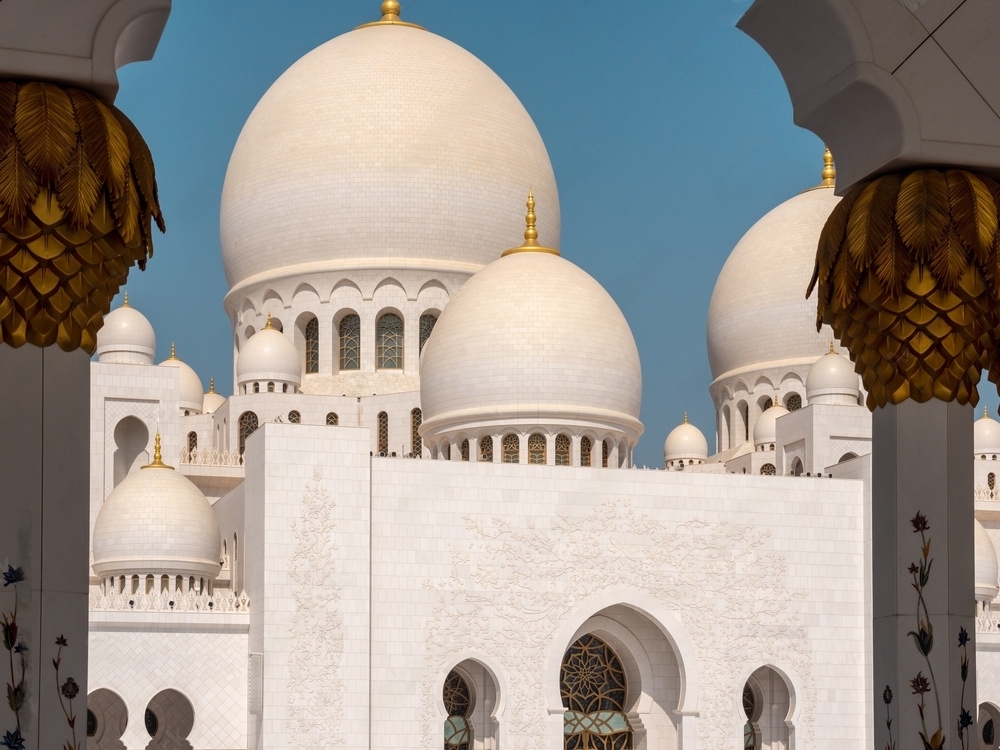} % Replace 'example-image-b' with the filename of your second image
    \caption{Religious Site}
    \label{fig:second}
  \end{minipage}\hfill
  \begin{minipage}{0.3\textwidth}
    \centering
    \includegraphics[width=\linewidth]{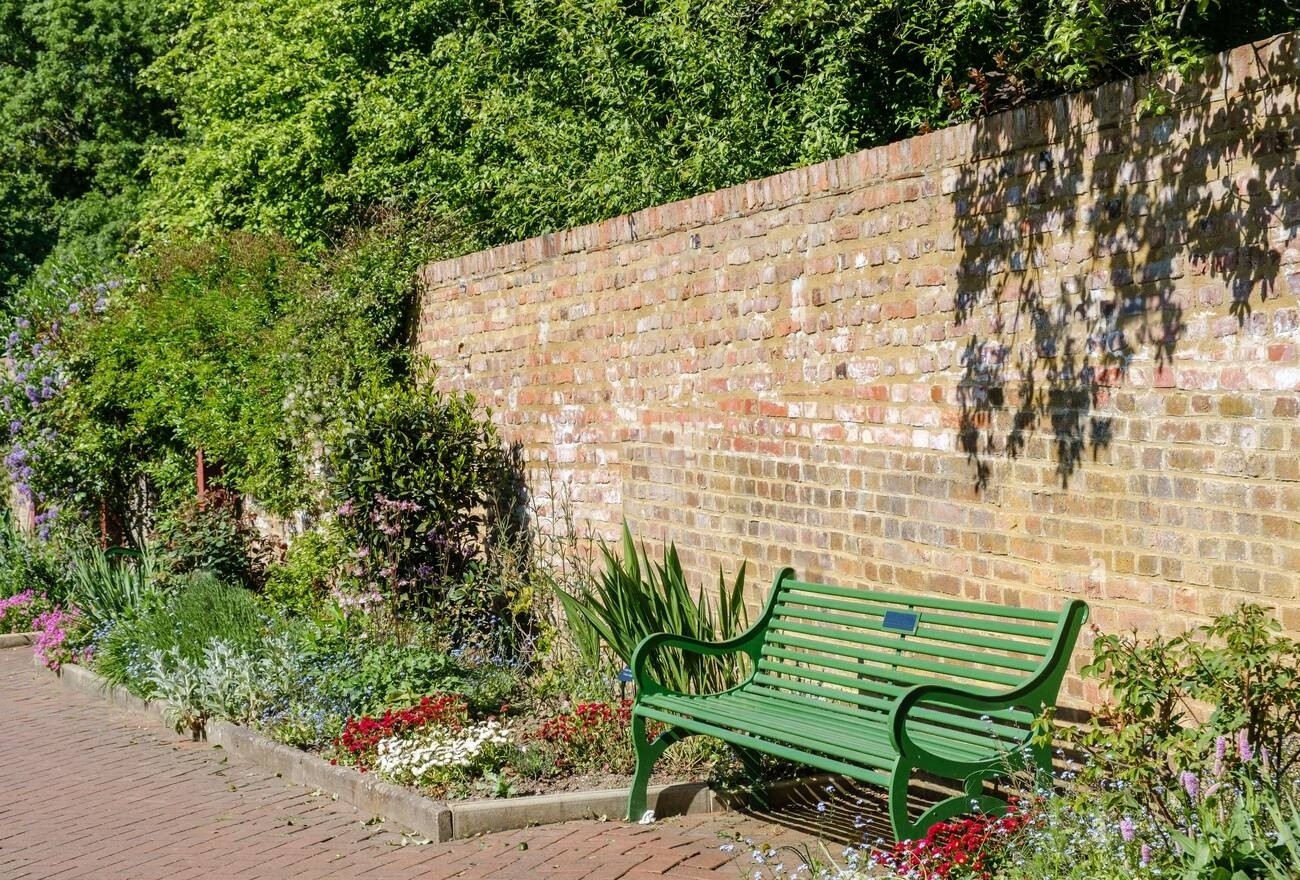} % Replace 'example-image-c' with the filename of your third image
    \caption{Public Space}
    \label{fig:third}
  \end{minipage}
\end{figure}
\begin{figure}[h]
  \centering
  \begin{minipage}{0.3\textwidth}
    \centering
    \includegraphics[width=\linewidth]{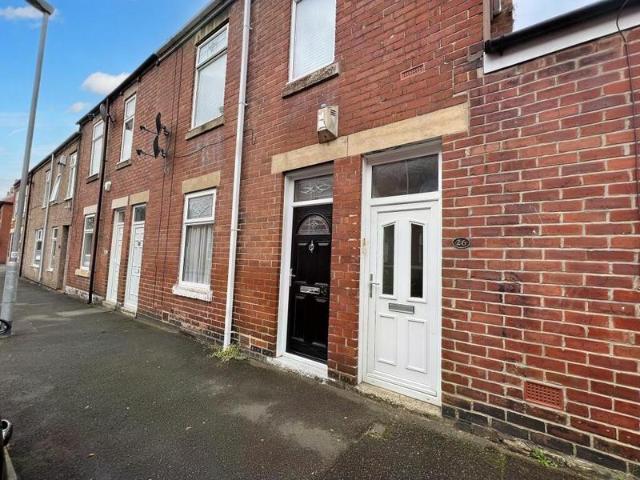} % Replace 'example-image-a' with the filename of your first image
    \caption{Residential Area}
    \label{fig:first}
  \end{minipage}\hfill
  \begin{minipage}{0.3\textwidth}
    \centering
    \includegraphics[width=\linewidth]{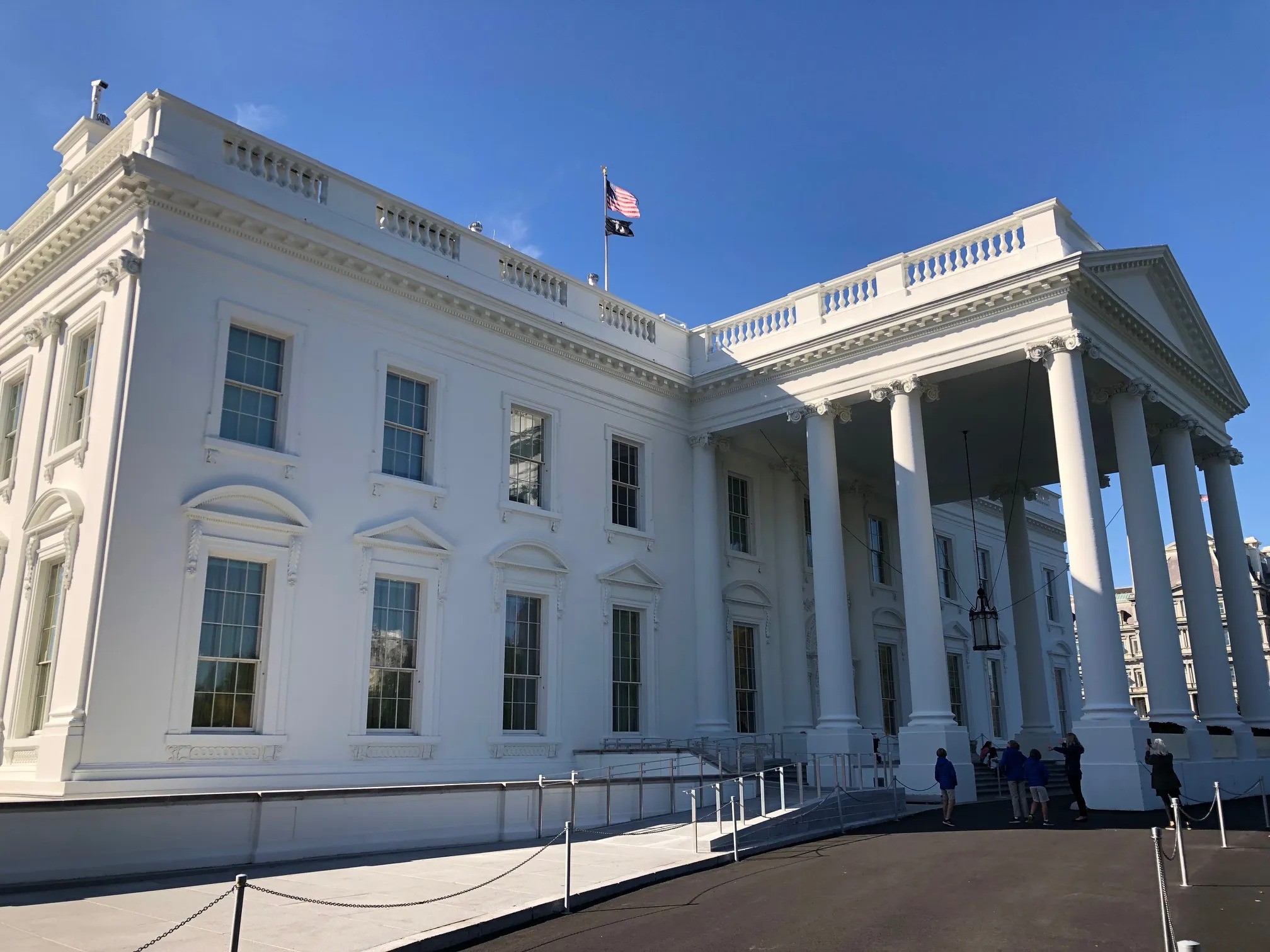} % Replace 'example-image-b' with the filename of your second image
    \caption{Government Building}
    \label{fig:second}
  \end{minipage}\hfill
  \begin{minipage}{0.3\textwidth}
    \centering
    \includegraphics[width=\linewidth]{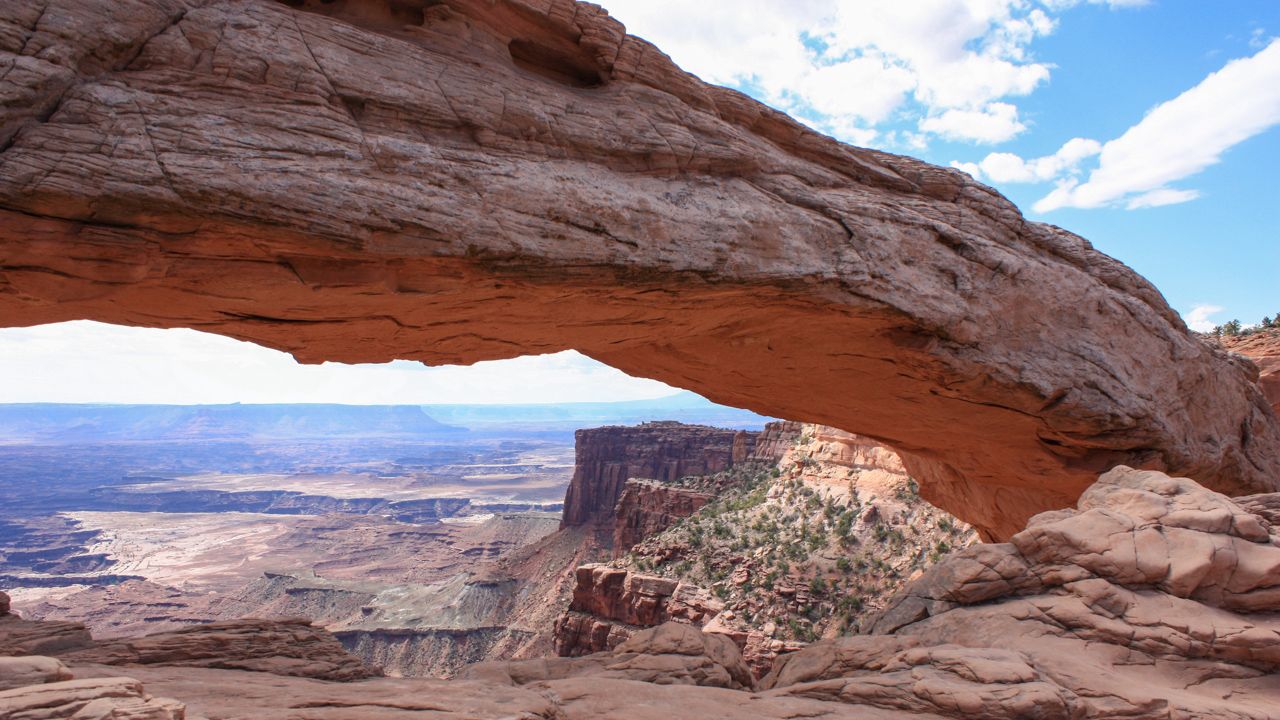} % Replace 'example-image-c' with the filename of your third image
    \caption{Tourist Point of Interest}
    \label{fig:third}
  \end{minipage}
\end{figure}

% \appendix

\end{document}